\documentstyle[preprint,prd,aps,psfig,floats,epsfig]{revtex}
\newcommand{\be}{\begin{equation}}
\newcommand{\ee}{\end{equation}}
\newcommand{\bea}{\begin{eqnarray}}
\newcommand{\eea}{\end{eqnarray}}
\renewcommand{\d}{{\rm d}}
\renewcommand{\L}{{\Lambda_{QCD}}}
\newcounter{dafigcounter}

\newcommand{\pfig}[3]{
 \refstepcounter{dafigcounter}
 \begin{minipage}[t]{#2}
  \begin{center}
   {\epsfxsize=#2 \mbox{\epsffile{#1.eps}}}
  \end{center}
  \label{#1}
  \small \bf Fig.~\thedafigcounter\rm\ #3
 \end{minipage}
}

\tightenlines
\begin{document}
\draft
\firstfigfalse
\title{Partonic Scattering Cross Sections in the QCD Medium}
\author{J\"org Ruppert, Gouranga C.~Nayak \thanks{present address: T-8, Theoretical Division, Los Alamos
National Laboratory, Los Alamos, NM 87545, USA}, Dennis D.~Dietrich, Horst St\"ocker,
        and Walter Greiner}
\address
{{\small\it{Institut f\"ur Theoretische Physik, J. W.
Goethe-Universit\"at, 60054 Frankfurt am Main, Germany}}}
\maketitle

\begin{abstract}
A medium modified gluon propagator is used to eva\-lu\-ate the
scattering cross section for the process $gg \rightarrow gg$ in
the QCD medium by performing an explicit sum over the
polarizations of the gluons. We incorporate a magnetic sreening
mass from a non - perturbative study.
 It is shown that the medium modified cross section
is finite, divergence free, and is independent of any ad-hoc
momentum transfer cut-off parameters. The medium modified finite
cross sections are necessary for a realistic investigation of the
production and equilibration of the minijet plasma expected at
RHIC and LHC.
\end{abstract}
\bigskip


A lot of research is undertaken to detect a new state of matter,
known as the quark-gluon plasma (QGP) which is believed to have
existed in the early stage of the universe, $\sim 10^{-4}$ seconds
after the big bang. \cite{qm2001,sqm2000} Lattice QCD calculations
indicate that this deconfined state of quark gluon matter exists
at high temperatures ($\sim 200~ MeV$) or high energy densities ($
\sim 2~ GeV/fm^{3}$) \cite{lattice}. The relativistic heavy-ion
colliders RHIC (Au-Au collisions at $\sqrt{s}=200~ A~GeV$) and LHC
(Pb-Pb collisions at $\sqrt{s}=5.5~ A~TeV$ ) offer the unique
opportunity to study the production of this state of matter in the
laboratory \cite{qm2001}. Perturbative QCD (pQCD) estimates that
the energy density of jets and minijets produced in these
collisions could be larger than about $50$ and $1000~ GeV/fm^{3}$
at RHIC and LHC respectively \cite{nayak,eskola}. However, even if
the attained energy density is sufficient to form a quark-gluon
plasma, it is not at all clear whether the parton medium will
equilibrate and form a thermalized quark-gluon plasma before
hadronization. Evidence for the formation of a QGP at RHIC and LHC
can only be established if various proposed signatures have been
examined experimentally. The most prominent signatures suggested
so far are $J/\Psi$ suppression \cite{Matsui}, strangeness
enhancement \cite{Rafelski},
 and dilepton and direct photon production
\cite{Strickland,Shuryak,Alam,GCN}. A quantitative calculation of
these predicted signatures will be difficult to achieve for the
situations found at RHIC and LHC. One main uncertainty originates
from the absence of a reliable study of the space-time evolution
of the parton distributions in these experiments. An investigation
of this evolution will determine the equilibration times and the
time evolution of all other local and global quantities such as
energy densities, number densities, and temperatures of the
SU(4)-flavour components of the quark-gluon plasma.

The space-time evolution of the QGP during an ultra-relativistic
heavy-ion collision might proceed through different stages such
as: 1) the pre-equilibrium, 2) the equilibrium, and 3) the
hadronization stage. During the equilibrium stage, hydrodynamics
can be used to describe the dynamics of the QGP. However, both the
pre-equilibrium and the hadronization stages have to be studied in
more detail in order to see whether either partons or hadrons
equilibrate or not. The first stage starts just after the initial
nuclear collisions, where many hard, semihard, and soft partons
are produced. The hard and semihard partons (jets and minijets)
formed at RHIC and LHC can be desribed by using pQCD
\cite{nayak,eskola,Wang}), and soft gluons may be described by the
creation of a coherent chromofield
\cite{bhal,larry,mueller,roberts,motolla}. For the sake of
simplicity, we restrict our investigation to jets and minijets
only. The equilibration of a minijet plasma can be studied by
solving the relativistic transport equations with binary and
secondary parton-parton collisions taken into account
\cite{geiger}: \be p^{\mu}\partial_{\mu}f(x,p)~=~C(x,p).
\label{trs1} \ee In equation (\ref{trs1}) \be C(x,p)~=~ \int
\frac{d^3p_2}{(2\pi)^3p_2^0}\frac{d^3p_3}{(2\pi)^3p_3^0}
\frac{d^3p_4}{(2\pi)^3p_4^0}~ |M(pp_2\rightarrow p_3p_4)|^2
~[f(x,p_3)f(x,p_4)-f(x,p)f(x,p_2)] \label{col1} \ee is the
collision term for the 2 body partonic scattering process $p
p_2\rightarrow p_3p_4$. The collisions among the partons determine
how the QGP reaches equilibrium. So, the investigation of the
2-body collision cross section in the non-equilibrated medium is
of great importance for the understanding of the minijet plasma
evolution at RHIC and LHC. The different partonic scattering
processes which have to be considered are e.g. $gg \rightarrow
gg$, $qq\rightarrow qq$, $q\bar q \rightarrow q \bar q$ etc. and
in more detailed studies also higher order processes like $gg
\rightarrow ggg$ might have to be taken into account. Most
probably, the dominant part of the minijets consists of gluons.
 Hence we consider collisions of the type:
$gg \rightarrow gg$. The differential cross section for this
process is given by: \be \frac{d \sigma}{d t} ~=~\frac{9\pi
\alpha_s^2}{2 s^2}~[3 -\frac{ut}{s^2} -\frac{u s}{t^2} -\frac{s
t}{u^2}], \label{gmat} \ee where $s=(q_1+q_2)^2=(q_3+q_4)^2$,
$t=(q_1-q_4)^2=(q_2-q_3)^2$, and $u=(q_1-q_3)^2=(q_2-q_4)^2$ are
the Mandelstam variables of the partons. For real gluons, they are
related by \bea t~=~-\frac{s}{2}~[1-\cos\theta_{\rm cm}] ~~~{\rm
and}~~~ u~=~-\frac{s}{2}~[1+\cos\theta_{\rm cm}], \label{tu} \eea
where $\theta_{\rm cm}$ is the center of mass scattering angle,
which goes from $0 \rightarrow \frac{\pi}{2}$ for identical
particles in the final state. If $|M(s,u,t)|^2$ is put from
Eq.(\ref{gmat}) into Eq.(\ref{col1}), a divergence emerges in the
collision term at small angle ($\theta_{\rm cm} \rightarrow$ 0),
which corresponds to collisions with small momentum transfer ($t
\rightarrow 0$). Hence, to obtain a finite collision term, one is
forced to put a minimum momentum transfer by hand, as it is done
in the parton cascade model \cite{pcm}. The plasma evolution and
the determination of all the signatures then crucially depend on
the choice of this momentum cut-off parameter.

The equilibration time and equilibration process can be studied by
using a collision term in the relaxation-time approximation for
the transport equation \cite{hosoya,zhang,nayak}:

\be
 C(x,p)~=~p^{\mu}u_{\mu}~[f(x,p)-f_{eq}(x,p)]/{\tau_c(\tau)},
\label{col2}
 \ee
where $\tau_c(\tau)$ is the time dependent relaxation time and
$f_{eq}=(e^{p^{\mu}u_{\mu}/T(\tau)}~-~1)^{-1}$ is the
Bose-Einstein distribution function. $u_{\mu}$ represents the flow velocity.
The time dependent relaxation time $\tau_c(\tau)$
can be determined from the transport cross
section ($\sigma_{tr}$) and the number density ($n$)
of the plasma via the relation
\cite{nayak,hosoya,daniel}:
 \be
\tau_c(\tau)~=~\frac{1}{\sigma_{tr}(\tau)~n(\tau)}.
 \label{reti}
 \ee
Here, the time dependent number density is determined from the
non-equilibrium distribution function via:
 \be n(\tau)~=~\int
\frac{d^3p}{(2\pi)^3p^0}~p^{\mu}u_{\mu}~f(x,p) \label{nd} \ee and
the transport cross section for the $gg \rightarrow gg$ collisions
is given by: \be \sigma_t(\tau)~=~\int_{-\frac{s}{2}}^0dt
~\frac{d\sigma^{2\rightarrow 2}}{dt}~\sin^2\theta_{\rm
cm}~=~\int_{-\frac{s}{2}}^0dt~\frac{1}{16\pi s^2}~
|M(s,t,u)|^2~\frac{4tu}{s^2}. \label{sgt} \ee Hence, whether one
uses Eq.(\ref{col1}) or (\ref{col2}) as the collision term, one
always encounters the divergence in the limit $t \rightarrow$ 0.

This divergence is inescapable as long as we are considering the
propagator in the vacuum. As the equilibration of the quark-gluon
plasma crucially depends on the cut-off used to remove these
divergences \cite{pcm}, it is essential to incorporate the medium
effects to obtain finite and cut-off independent partonic
scattering cross-sections in the QCD medium. Here, one should
therefore use the the medium modified propagator instead of the
vacuum propagator. In this paper, we evaluate and analyze the
matrix element squared, total cross section, and transport cross
section for the process $gg \rightarrow gg $ by using the medium
modified propagator performing an explicit sum over the
polarizations of the physical gluons. This analysis is important
to study the evolution of the QGP because the divergence is
removed automatically and cut-off independently.

In this paper we choose the medium modified gluon propagator in
the covariant gauge. It can be split into the longitudinal and
transverse parts \cite{weldon,landshoff}: \bea
\Delta^{\mu\nu}(K)=\frac{P_{T} ^{\mu \nu}}{-K^2 + \Pi_T}
+\frac{P_{L}^{\mu \nu}}{-K^2 +\Pi_L} +(\alpha-1)
\frac{K^{\mu}K^{\nu}}{K^4} \label{prop} \eea where $\alpha$ is the
gauge fixing parameter. $P^{\mu\nu}_{T}$ and $P^{\mu\nu}_{L}$ are
longitudinal and transverse tensors given by: \bea
P_{L}^{\mu \nu}&=&\frac{-(\omega K^\mu - K^2 U^\mu)(\omega K^\nu - K^2 U^\nu)}{K^2 k^2} \nonumber \\
P_{T}^{\mu \nu}&=&\tilde{g}^{\mu \nu}+ \frac{\tilde{K}^{\mu}
\tilde{K}^{\mu}}{k^2}, \eea $\tilde{g}_{\mu\nu}=g_{\mu \nu}-U_\mu
U_\nu$ with $U_{\mu}$ being the flow velocity. The quantity
$\omega=K \cdot U$ is interpreted as the Lorentz-invariant energy
and $k=\sqrt{-\tilde{K}_\mu \tilde{K}^\mu}$ with
$\tilde{K}_\mu~=~K_\mu - U_\mu(K \cdot U)$ as the three momentum
of the virtual boson. In the local rest frame ($U=(1,\vec{0})$),
$\omega$ and $\vec{k}$ are its energy and momentum. The
expressions for $\Pi_L$ and $\Pi_T$ in the high temperature
expansion are derived by perturbative methods in \cite{weldon}:
\bea \Pi_L&=&m_D^2 (1-x^2) [1-\frac{x}{2}
\log{|\frac{1+x}{1-x}|}+i \frac{\pi}{2}x]
\nonumber \\
\Pi_T&=&m_D^2
[\frac{x^2}{2}+\frac{x}{4}(1-x^2)\log{|\frac{1+x}{1-x}|}
-i\frac{\pi}{4}x(1-x^2)], \eea where $x=\frac{\omega}{k}$ and
$m_D=g^2 T^2$. The t-channel matrix element for the process  $gg
\rightarrow gg$ in the medium is given by: \bea M =  g^2 f_{aed}
f_{ebc}
     \epsilon_1^\lambda \epsilon_4^{\sigma}
     V^{\lambda \tau \sigma}(-q_1,q_1-q_4,q_4) \Delta_{\tau ' \tau}(q_1-q_4)
     V^{\tau ' \mu \nu}(q_2-q_3,-q_2,q_3)
     \epsilon_2^\mu \epsilon_3^\nu,
\eea
where $V^{\mu \lambda \nu}(p_1,p_2,p_3)=[(p_1-p_2)^\nu g^{\mu \lambda}+
                                         (p_2-p_3)^\mu g^{\lambda \nu}+
                                         (p_3-p_1)^\lambda g^{\mu \nu}]$
is the three gluon vertex, $\epsilon^{\mu}(p)$ are the polarization
vectors of the gluons and $\Delta^{\mu \nu}(k)$ is the medium modified
gluon propagator given by Eq. (\ref{prop}).

To sum over initial and final spins of the gluons in the matrix element
squared, we use the appropriate projection operators for the transverse
polarization states of the gluons \cite{cutler,combridge}: \bea
\sum_{\rm spins} \epsilon_1^\lambda \epsilon_1^{*\lambda'} = -
g^{\lambda \lambda'}+\frac{2(q^\lambda_1 q^{\lambda'}_2
+q^{\lambda'}_1 q^{\lambda}_2 )}{(q_1+q_2)^2} &~~,~~~~~& \sum_{\rm
spins} \epsilon_2^\mu \epsilon_2^{*\mu '} = - g^{\mu
\mu'}+\frac{2(q^\mu_1 q^{\mu'}_2 +q^{\mu'}_1 q^{\mu}_2
)}{(q_1+q_2)^2}
\nonumber \\
\sum_{\rm spins}  \epsilon_3^\nu \epsilon_3^{*\nu'} = - g^{\nu \nu'}+\frac{2(q^\nu_3 q^{\nu'}_4 +q^{\nu'}_3 q^{\nu}_4 )}{(q_3+q_4)^2} &~~,~{\rm and}~& \sum_{\rm spins}  \epsilon_4^\sigma \epsilon_4^{*\sigma'} = - g^{\sigma \sigma'}.
\eea
After summing over the final and averaging over the initial
spins and color, we obtain (in Feynman gauge):
\bea
|M|^2_{\rm medium} =    \frac{A}{s^3 (t-(t_{\rm v} \cdot
U)^2)^2 (\Pi_L-t)(\bar{\Pi}_L-t)}
  - \frac{B}{s^3 (t-(t_{\rm v} \cdot U)^2)^2 (\Pi_T-t) (\bar{\Pi}_L-t)}
\nonumber \\
  - \frac{B}{s^3 (t-(t_{\rm v} \cdot U)^2)^2 (\Pi_L-t)(\bar{\Pi}_T-t)}
  + \frac{C}{s^3 (t-(t_{\rm v} \cdot U)^2)^2 (\Pi_T-t)(\bar{\Pi}_T-t)},
\label{mfull} \eea where $A$, $B$, $C$ are: \bea A &=& \frac{9}{8}
t^2 g^4 \{ (s^2~ [(s_{\rm v} \cdot U - u_{\rm v} \cdot U)^2 + 4
(t-(t_{\rm v} \cdot U)^2)] +2ts~  [2t + (u_{\rm v} \cdot
U)^2-(s_{\rm v} \cdot U)^2 - 2 (t_{\rm v} \cdot U)^2  ]+
\nonumber \\
&& 2t^2~ [(u_{\rm v} \cdot U)^2+2(u_{\rm v} \cdot U) (s_{\rm v}
\cdot U) - (s_{\rm v} \cdot U)^2] ) \times (t~ [2t+ ( u_{\rm v}
\cdot U)^2 -2 (t_{\rm v} \cdot U)^2-(s_{\rm v} \cdot U)^2]
\nonumber \\
&& {}  +s~ [(s_{\rm v} \cdot U + u_{\rm v} \cdot U)^2+3(t-(t_{\rm
v} \cdot U)^2)])\} \label{A} \eea
\bea B &=& \frac{9}{8} t g^4 \{
2 s^4~ [t-(t_{\rm v} \cdot U)^2] [(u_{\rm v} \cdot U)^2-(s_{\rm v}
\cdot U)^2]
  + s^3 t ~ [(s_{\rm v} \cdot U)^4+ 2(t-(t_{\rm v} \cdot
U)^2-(u_{\rm v} \cdot U)^2)
\nonumber \\
&&
 (s_{\rm v} \cdot U)^2
 +2 (t-(t_{\rm v} \cdot U)^2) (u_{\rm v} \cdot U)
(s_{\rm v} \cdot U)
   +  (u_{\rm v} \cdot U)^2 (4t+(u_{\rm v} \cdot U)^2-4(t_{\rm v} \cdot U)^2)]
\nonumber \\
&& +s^2 t^2 [-3 (s_{\rm v} \cdot U)^4 - 2 (u_{\rm v} \cdot
U)(s_{\rm v} \cdot U)^3
   +  2(t-(t_{\rm v} \cdot U)^2)(s_{\rm v} \cdot U)^2
   +  (u_{\rm v} \cdot U)(-5(t_{\rm v} \cdot U)^2
\nonumber \\
&&+ 2 (u_{\rm v} \cdot U)^2+5 t) (s_{\rm v} \cdot U)
+ (u_{\rm v}
\cdot U)^2(-5(t_{\rm v} \cdot U)^2+3(u_{\rm v} \cdot U)^2+5t)]
 +2 s t^3  (u_{\rm v} \cdot U)
\nonumber \\
&& [2 (u_{\rm v} \cdot U)(s_{\rm v} \cdot U)^2+4(-(t_{\rm v} \cdot
U)^2 +(u_{\rm v} \cdot U)^2+t)(s_{\rm v} \cdot U) + (u_{\rm v}
\cdot U)(-3(t_{\rm v} \cdot U)^2 +
\nonumber \\
&&2(u_{\rm v} \cdot U)^2+3t)]
 +2 t^4
   ~[s_{\rm v} \cdot U+u_{\rm v} \cdot U]
\times [(s_{\rm v} \cdot U)^3-3 u_{\rm v} \cdot U (s_{\rm v} \cdot U)^2
\nonumber \\
 &&
+ (u_{\rm v} \cdot U)^2 (s_{\rm v} \cdot U)+
(u_{\rm v} \cdot U) (t+ (u_{\rm v} \cdot U)^2-(t_{\rm v} \cdot
U)^2)] \} \label{B} \eea

\bea C &=&\frac{9}{8} g^4 \{ 4 s^5 ~[t- (t_{\rm v} \cdot U)^2]^2 +
4 s^4 t~ [t - (t_{\rm v} \cdot U)^2]~[t-(s_{\rm v} \cdot U)^2
-(t_{\rm v} \cdot U)^2+(u_{\rm v} \cdot U)^2]
\nonumber \\
&& {} + s^3 t^2~[(s_{\rm v} \cdot U)^4+(3(t_{\rm v} \cdot U)^2- 2
(u_{\rm v} \cdot U)^2-3t)
 (s_{\rm v} \cdot U)^2+2((t-(t_{\rm v} \cdot U)^2) (u_{\rm v} \cdot U)(s_{\rm v} \cdot U)
\nonumber \\
&& (u_{\rm v} \cdot U)^4 + 4(t- (t_{\rm v} \cdot U)^2)^2 +
(t-(t_{\rm v} \cdot U)^2) (u_{\rm v} \cdot U)^2)] + s^2 t^3 ~ [-3
(s_{\rm v} \cdot U)^4
\nonumber \\
&&- 2 (u_{\rm v} \cdot U)(s_{\rm v} \cdot U)^3 + 8 (t - (t_{\rm v}
\cdot U)^2)(s_{\rm v} \cdot U)^2 +
 2 u_{\rm v} \cdot U ( - 3 (t_{\rm v} \cdot U)^2 +
(u_{\rm v} \cdot U)^2+ 3t) s_{\rm v} \cdot U
\nonumber \\
&& + 3 (u_{\rm v} \cdot U)^4 + 7 (t - (t_{\rm v} \cdot U)^2)^2 -
6(t-t_{\rm v} \cdot U ^2) (u_{\rm v} \cdot U)^2)] + 2 s t^4 ~[2
(u_{\rm v} \cdot U)^4 + 4(s_{\rm v} \cdot U) (u_{\rm v} \cdot U)^3
\nonumber \\
&& + (2 (s_{\rm v} \cdot U)^2 + (t_{\rm v} \cdot
U)^2-t)(u_{\rm v} \cdot U)^2
 +2 s_{\rm v} \cdot U (t-(t_{\rm v} \cdot U)^2) u_{\rm
v} \cdot U+ (t-(t_{\rm v} \cdot U)^2)(7(s_{\rm v} \cdot
U)^2
\nonumber \\
&& -(t_{\rm v} \cdot U)^2+t)] + 2 t^5~[(s_{\rm v} \cdot U)^4-2
u_{\rm v} \cdot U (s_{\rm v} \cdot U)^3 +2(-(t_{\rm v} \cdot
U)^2-(u_{\rm v} \cdot U)^2+t)(s_{\rm v} \cdot U)^2
\nonumber \\
&& +2u_{\rm v} \cdot U ((t_{\rm v} \cdot U)^2+(u_{\rm v} \cdot
U)^2-t)s_{\rm v} \cdot U +(u_{\rm v} \cdot U)^4-(t-(t_{\rm v}
\cdot U)^2)^2] \}. \label{C} \eea
 $U$ is the flow velocity and $s_{\rm v},~t_{\rm v},~u_{\rm v}$ are
defined as follows \bea s_{\rm
v}^{\mu}=(q_1+q_2)^{\mu}=(q_3+q_4)^{\mu},~ t_{\rm
v}^{\mu}=(q_3-q_2)^{\mu}=(q_1-q_4)^{\mu},~ u_{\rm
v}^{\mu}=(q_1-q_3)^{\mu}=(q_4-q_2)^{\mu}. \eea The Mandelstam
variables of the partons are just given as: \bea s_{v} \cdot
s_{v}=s,~t_{v} \cdot t_{v}=t,~u_{v} \cdot u_{v}=u.\eea Eq.
(\ref{mfull}) together with Eqs. (\ref{A}) to (\ref{C}) show the
general expression for the t-channel matrix element squared of the
scattering process $gg \rightarrow gg$ in the medium. As this is a
rather lengthy expression, we analyze the behavior of this
scattering term in the local rest frame of the fluid
$U=(1,\vec{0})$. In the center of mass frame of the parton-parton
collision Eq. (\ref{mfull}) becomes:

\bea && {} (\frac{d \sigma}{dt})_{\rm medium}= \frac{9 \pi
\alpha^2}{8} \{
   \frac{(s^2 +2 t s +2 t^2)^2}{s^4(\Pi_L-t)(\bar{\Pi}_L-t)}
 + \frac{(s+t)(s^2-2t^2)}{s^3(\Pi_T-t)(\bar{\Pi}_L-t)}
\nonumber \\
&& {} + \frac{(s+t)(s^2-2t^2)}{s^3(\Pi_L-t)(\bar{\Pi}_T-t)}
 + \frac{(s+t)(s^4-3ts^3+15t^2s^2+8t^3s-2t^4)}{s^5(\Pi_T-t)(\bar{\Pi}_T-t)}
 \}.
 \label{dsdtmedium}
\eea Using the in-vacuum gluon propagator and repeating the
above procedure one finds for the t-channel process: \bea
  (\frac{d \sigma}{dt})_{\rm vacuum}=\frac{9 \pi \alpha^2}{8}
  \frac{4s^5+4ts^4+16t^2s^3+27t^3s^2+10t^4s-2t^5}{s^5 t^2}.
  \label{tfree}
\eea
It can be verified that for $\Pi_L = 0$ and $\Pi_T =0$
Eq.(\ref{dsdtmedium}) reduces to Eq.(\ref{tfree}).
In the limit $t \rightarrow 0$ the dominant divergent part of
Eq.(\ref{tfree}) is given by:

\bea
    (\frac{d \sigma}{dt})_{\rm lead. vacuum}
= \frac{9 \pi \alpha^2}{2t^2}.
\eea

This divergence can be removed by introducing a screening mass $m_D$ by
hand \cite{zhang,nayak}:

\bea
    (\frac{d \sigma}{dt})_{\rm lead. cut-off}
     = \frac{9 \pi \alpha^2}{2(t-m_D^2)^2}.
\label{vacuumT}
\eea
The introduction of the Debye screening mass (the longitudinal part of the
self energy) in the above formula
leads to a screening of the long range electric field but not of
the magnetic field.
However, Eq. (\ref{dsdtmedium}) can be used to screen both the electric
and magnetic (beyond one loop in self energy evaluation)
part simultaneously. For a comparison with Eq. (\ref{vacuumT})
we extract the terms quadratic in $t$ in the denominator of Eq. (\ref{dsdtmedium}):
\bea
 (\frac{d\sigma}{dt})_{\rm lead. medium}
= \frac{9 \pi \alpha^2}{8} \{ \frac{1}
{(\Pi_L-t)(\bar{\Pi}_L-t)}
 + \frac{1}{(\Pi_T-t)(\bar{\Pi}_L-t)}
 \nonumber \\
 + \frac{1}{(\Pi_L-t)(\bar{\Pi}_T-t)}
 +\frac{1}{(\Pi_L-t)(\bar{\Pi}_T-t)}\}.
\label{dsdtmediumT} \eea This equation gives the principal
contribution to the total cross section (see below). Hence, this
expression can be used for all practical purposes during the
evolution of the plasma. In the situation we are considering, it
can be easily checked that in the leading order $\Pi_L=m_D^2=g^2
T^2$ and $\Pi_T=0$. This suggests that even if we consider
Eq.(\ref{dsdtmediumT}), the magnetic part is still not screened
as long as we use the perturbative expression of the self-energies
in the leading order. In this study, the logarithmic singularity in the
transverse part of the total cross section is removed by
introducing a non-perturbative magnetic screening mass $m_{\rm
mag}^2= \frac{3}{2} (0.255 g^2 T)^2$ taken from
\cite{biro,heiselberg}. With these self energies
Eq.(\ref{dsdtmediumT}) can be integrated to obtain the total cross
section: \bea
 (\sigma_{\rm tot})_{\rm lead. medium}&=&~\int_{-\frac{s}{2}}^0dt
~(\frac{d\sigma}{dt})_{\rm lead. medium}
\nonumber \\
&=& \frac{9 \pi \alpha^2 }{8}(\frac{s}{\Pi_L (s+2
\Pi_L)}+\frac{1}{\Pi_T}+
 2\frac{ \ln(\frac{s}{\Pi_L}+2)-\ln(\frac{s}{\Pi_T}+2)}{\Pi_T-\Pi_L}
 -\frac{2}{s+2 \Pi_T}),
\label{totmedium}
\eea
and the transport cross section:
 \bea &&(\sigma_{\rm tr})_{\rm lead. medium}=\int_{-\frac{s}{2}}^0dt
~(\frac{d\sigma}{dt})_{\rm lead. medium}~\frac{4ut}{s^2} =
\nonumber
\\
 &&=~ \frac{9 \pi \alpha^2 }{4}\{2\frac{\ln(\frac{s}{2 \Pi_L} +
1)+\ln(\frac{s}{2 \Pi_T} + 1)}{s}
-\frac{4}{s}-\frac{1}{s+2\Pi_L}-\frac{1}{s+2\Pi_T}
 \nonumber \\
  && {}+
\frac{
8 \ln(\frac{s}{2 \Pi_L} + 1)\Pi_L^2 -2(s(1-2\ln(\frac{s}{2 \Pi_L }+1))
+2(\ln(\frac{s}{2 \Pi_L }+1)-\ln(\frac{s}{2 \Pi_T}+1))\Pi_T)\Pi_L }
{s^2(\Pi_L-\Pi_T)} \nonumber \\
&& {} +\frac{2\Pi_T(s(1-2\ln(\frac{s}{2\Pi_T}+1))-4\ln(\frac{s}{2 \Pi_T
}+1)\Pi_T)}{s^2(\Pi_L-\Pi_T)}\}.
\label{transmedium}
\eea

To show that the above leading order (in t) cross sections give the
dominant contribution, we evaluate the total and transport cross
sections by using the t-channel expression for
$\frac{d\sigma}{dt}$ as given by  Eq.(\ref{dsdtmedium}) and then
plot the differences between the full and the leading order
expressions for the total cross section  in Fig.1. The graphs show
that the difference between the full cross sections and the
leading order cross sections are very small. So, for practical
purposes, one can use Eq. (\ref{totmedium}) and
(\ref{transmedium}) as the total and transport cross sections for
the process $gg \rightarrow gg$ in the medium. In the figures we
use $T=\frac{\sqrt{s}}{5.4}$ and $\alpha=0.3$.

The importance of the medium modified total and transport cross
sections given in Eq. (\ref{totmedium}) and (\ref{transmedium})
will be discussed below. To demonstrate it, we evaluate the total
and transport cross section by using Eq. (\ref{vacuumT}) which
yields: \bea (\sigma_{\rm tot})_{\rm lead.
cut-off}~=~\int_{-\frac{s}{2}}^0dt ~(\frac{d\sigma}{dt})_{\rm
lead. cut-off} ~= ~ (\frac{9 \pi \alpha^2 s }{4 \Pi_L^2+2s\Pi_L}),
\label{totold} \eea and \bea
 (\sigma_{\rm tr})_{\rm lead. cut-off}
 &=&\int_{-\frac{s}{2}}^0dt ~(\frac{d\sigma}{dt})_{\rm lead.
 cut-off} ~\frac{4ut}{s^2}
\nonumber \\
 &=& ~ 9 \pi \alpha^2
 (2 \frac{\ln (\frac{s}{2 \Pi_L}+1)-1}{s}+
               4 \frac{\Pi_L \ln (\frac{s}{2 \Pi_L}+1)}{s^2}
               -\frac{1}{s+2 \Pi_L}).
\label{transold} \eea These are the cross sections obtained by
introducing a Debye screening mass by hand into the vacuum formula
(see Eq. (\ref{vacuumT})). As already mentioned, there is no way
that magnetic screening is incorporated in these formulas.
However, one could argue that the contribution from the magnetic
sector to the total and transport cross section is small. This is
to be checked here. Since any significant change in the total and
transport cross sections crucially influences the predictions for
all the global quantities and the signatures of the quark-gluon
plasma (see \cite{nayak}), it is important to compare our medium
modified  cross section (both electric and magnetic screening
taken into account) with that of the Debye screened cross sections
obtained by using Eq. (\ref{vacuumT}). In Fig. \ref{sigmatot} we
plot the total cross section obtained by using the medium modified
propagator including electric and magnetic sreening (see Eq.
(\ref{totmedium})) and  the total cross section obtained by using
the in-vacuum propagator with a Debye screening mass as a cut-off
(see Eq. (\ref{totold})). In this figure, we use the values of $s$
and $\alpha$ which correspond to a "realistic" situation at RHIC
\cite{nayak}. It can be seen that  for the values of $s$ and
$\alpha$ considered here, the difference in the two cross sections
increases for decreasing values of $s$.
 In Fig.\ref{sigmatrans}, we plot the
transport cross sections for the two cases described above
(Eq.(\ref{transmedium}) and (\ref{transold})). The transport cross
section directly enters into the collision term (see Eq.
\ref{col2}) and so determines the evolution of the quark-gluon
plasma. Similar to the total cross section, it can be seen that
the medium modified transport cross section is enhanced in
comparison to the other for smaller values of $s$. Hence, it can
be expected that all the global quantities and the equilibration
time might change when the vacuum propagator is replaced by the
medium modified one. However, this has to be checked by including
all these features in a self consistent transport study. For
example, in the above plots, for simplicity we have used a
constant value of  $\alpha =0.3$. In "realistic" situations at
RHIC and LHC the $\alpha$, $s$, total cross section, transport
cross section, as well as all other quantities are time dependent
and, therefore, have to be calculated via the distribution
functions of the partons, which in turn are obtained by solving
the transport equation self-consistently \cite{nayak}. As the
self-consistent transport study involves extensive additional
numerical work,  the results incorporating the medium modified
transport cross section will be presented elsewhere.

In summary, the medium modified propagator has been used to
evaluate the partonic scattering cross section for the process $gg
\rightarrow gg$ in the QCD medium by performing an explicit sum
over the physical gluon polarizations. A magnetic screening mass
from a non-perturbative study was used to show that the medium
modified cross section is finite, divergence free, and is uniquely
determined. The medium modified cross sections yield different
results than these obtained by artificially introducing a Debye
screening mass by hand into the vacuum formulae. This implies that
the medium modified scattering parton cross sections must be
incorporated properly into the transport equations in order to
study the production and possible equilibration of the minijet
plasma at RHIC and LHC. Any change of the cross sections due to
in-medium effects will crucially change the equilibration times,
time evolution of the energy densities, number densities,
temperatures of all degrees of freedom and hence, all predictions
of signatures of the quark-gluon plasma, both at RHIC and LHC.

\section*{Acknowledgements}
We thank Dirk H. Rischke, Stefan Hofmann, and Chung-Wen Kao
for helpful discussions. G.C.N. would like to thank the
Alexander von Humboldt Foundation and the
BMBF for financial support. D.D.D. would like to thank the
Graduiertenf\"orderung des Landes Hessen for financial support.

\begin{figure}[thb]
\begin{center}
\pfig{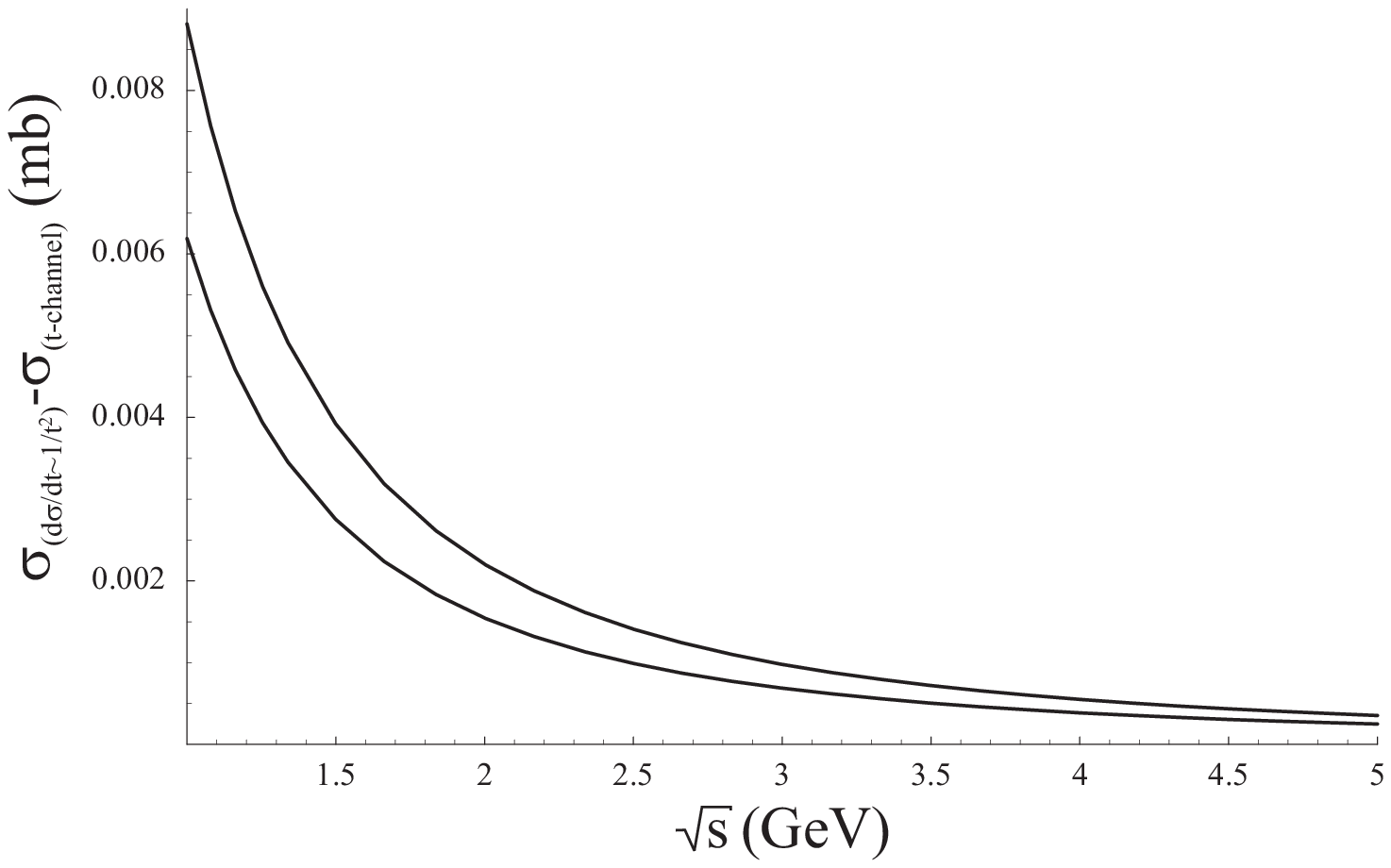}{12cm}
 {In the upper curve the difference of the total cross
  section in small angle approximation to
  the same quantity obtained by using the t-channel matrix element
 $|\sigma_{(\frac{d\sigma}{dt} \propto \frac{1}{t^2})}-\sigma_{\rm
 (t-channel)}|$is plotted as a function of $\sqrt{s}$ ($\alpha = 0.3$).
 The lower curve is the difference obtained  for the transport cross sections.
 }
\end{center}
\end{figure}

\begin{figure}[thb]
\begin{center}
\pfig{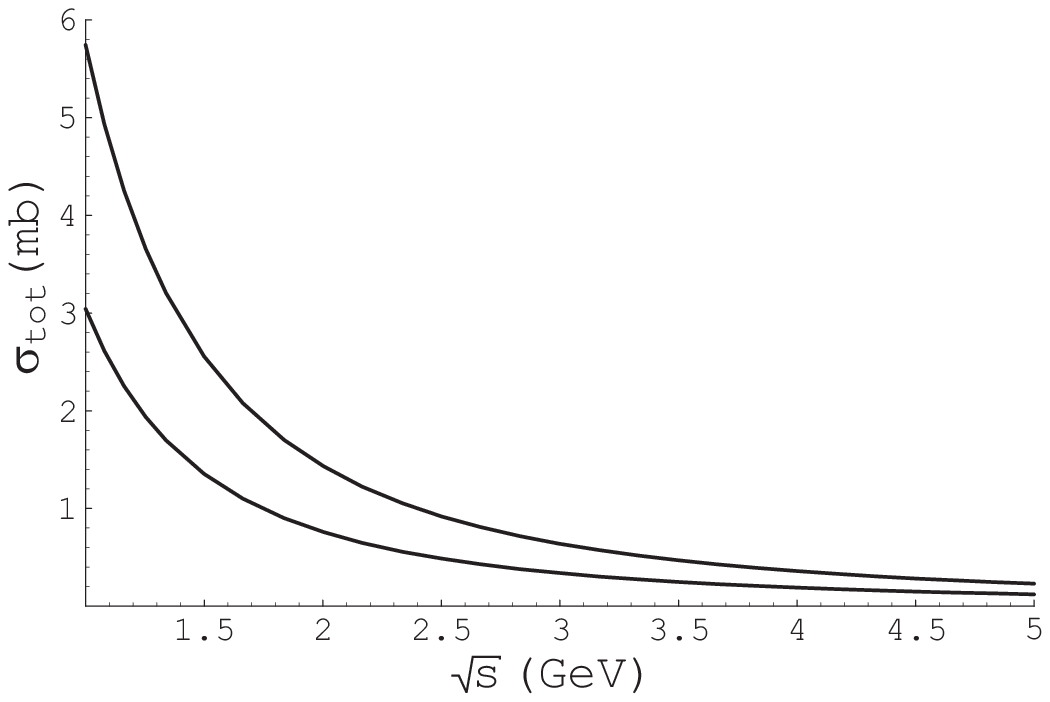}{12cm}
 {The total in-medium cross section  and
  the total cross section regularized by putting a cut-off by hand
 $\sigma_{\rm tot}$
 is plotted as a function of $\sqrt{s}$. ($\alpha = 0.3$)
 The first is significantly enhanced in comparison to
 the second.}
\end{center}
\end{figure}

\begin{figure}[thb]
\begin{center}
\pfig{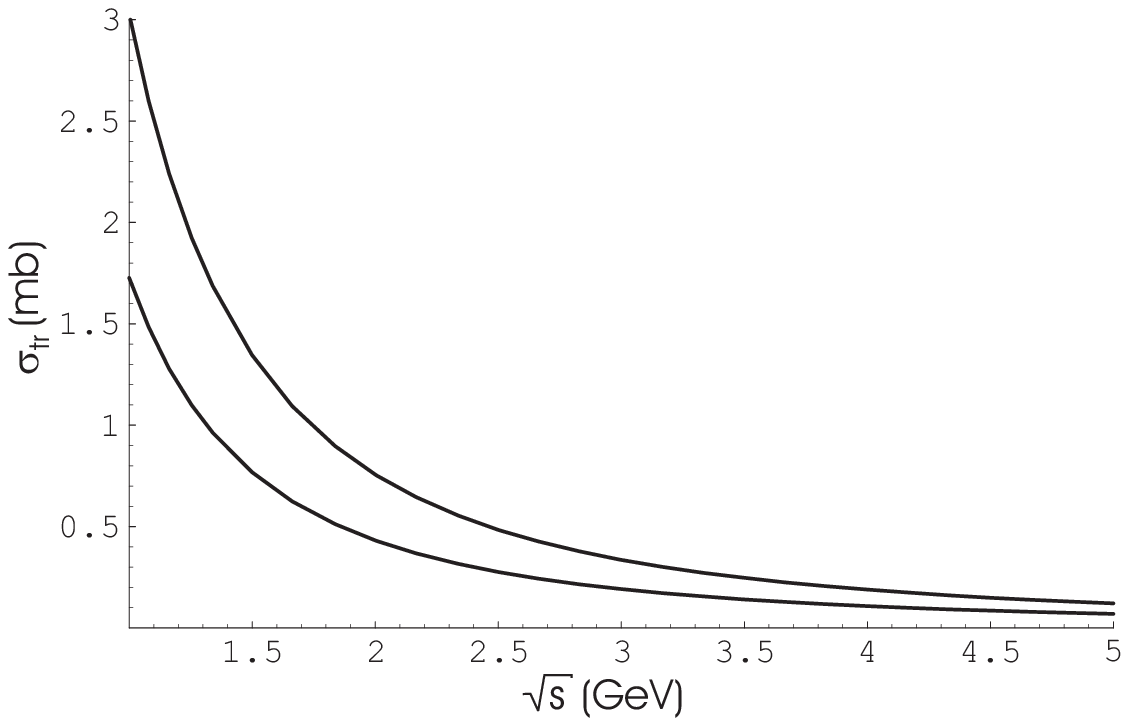}{12cm}
 {The in-medium transport cross section  and
  the transport cross section regularized by putting a cut-off by hand
 $\sigma_{\rm trans}$
 is plotted as a function of $\sqrt{s}$.($\alpha = 0.3$)
 The first is significantly enhanced in comparison to
 the second.}
\end{center}
\end{figure}

\begin{thebibliography}{99}

\bibitem{qm2001}
Quark matter 2001. Proceedings, 15th International Conference on
ultra-relativistic nucleus nucleus collisions, QM 2001, Stony
Brook, January 2001.

\bibitem{sqm2000}
Strangeness in Quark Matter 2000, 5th International Conference on
Strangeness in Quark Matter, Berkeley, California, July 2000.

\bibitem{lattice}
See, {\it e.g.}

L.~D.~McLerran and B.~Svetitsky,
Phys.\ Rev.\ D {\bf 24}, 450 (1981);
L.~McLerran,
Phys.\ Rev.\ D {\bf 36}, 3291 (1987);
R.V. Gavai, in {\it Quantum Fields on the Computer},
ed. M. Creutz, (World Scientific,
1992), p. 51;
F.~Karsch and E.~Laermann,
Rept.\ Prog.\ Phys.\  {\bf 56}, 1347 (1993)
[hep-lat/9304010];
M.~Oevers, F.~Karsch, E.~Laermann and P.~Schmidt,
Nucl.\ Phys.\ Proc.\ Suppl.\  {\bf 63}, 394 (1998)
[hep-lat/9709071],
Nucl.\ Phys.\ B {\bf 605}, 579 (2001) [hep-lat/0012023].


\bibitem{nayak}

G.~C.~Nayak, A.~Dumitru, L.~McLerran and W.~Greiner,
Nucl.\ Phys.\ A {\bf 687}, 457 (2001)
[hep-ph/0001202].

\bibitem{eskola}
K.~J.~Eskola and K.~Kajantie,
Z.\ Phys.\ C {\bf 75}, 515 (1997)
[nucl-th/9610015];
N.~Hammon, H.~Stocker and W.~Greiner,
Phys.\ Rev.\ C {\bf 61}, 014901 (2000)
[hep-ph/9903527].

\bibitem{Matsui}
T.~Matsui and H.~Satz,
Phys.\ Lett.\ B {\bf 178}, 416 (1986).
\bibitem{Rafelski}
J.~Rafelski and B.~Muller,
Phys.\ Rev.\ Lett.\  {\bf 48}, 1066 (1982)
[Erratum-ibid.\  {\bf 56}, 2334 (1982)].

\bibitem{Strickland}
M.~Strickland,
Phys.\ Lett.\ B {\bf 331}, 245 (1994).

\bibitem{Shuryak}
E.~Shuryak and L.~Xiong,
Phys.\ Rev.\ Lett.\  {\bf 70}, 2241 (1993)
[hep-ph/9301218].


\bibitem{Alam}

J.~Alam, B.~Sinha and S.~Raha,
Phys.\ Rept.\  {\bf 273}, 243 (1996).


\bibitem{GCN}
G.~C.~Nayak,
Phys.\ Lett.\ B {\bf 442}, 427 (1998)
[hep-ph/9801321].


\bibitem{Wang}
X.~Wang and M.~Gyulassy,
Phys.\ Rev.\ D {\bf 44}, 3501 (1991);
X.~Wang,
Phys.\ Rept.\  {\bf 280}, 287 (1997)
[hep-ph/9605214].

\bibitem{bhal}

R.~S.~Bhalerao and G.~C.~Nayak,
Phys.\ Rev.\ C {\bf 61}, 054907 (2000)
[hep-ph/9907322].

\bibitem{larry}


L.~McLerran and R.~Venugopalan,
Phys.\ Rev.\ D {\bf 49}, 2233 (1994)
[hep-ph/9309289].

\bibitem{mueller}
Y.~V.~Kovchegov and A.~H.~Mueller,
Nucl.\ Phys.\ B {\bf 529}, 451 (1998)
[hep-ph/9802440];
A.~H.~Mueller,
Nucl.\ Phys.\ B {\bf 572}, 227 (2000)
[hep-ph/9906322].

\bibitem{roberts}
C.~D.~Roberts and S.~M.~Schmidt,
Prog.\ Part.\ Nucl.\ Phys.\  {\bf 45S1}, 1 (2000)
[nucl-th/0005064].

\bibitem{motolla}
Y.~Kluger, J.~M.~Eisenberg, B.~Svetitsky, F.~Cooper and E.~Mottola,
Phys.\ Rev.\ Lett.\  {\bf 67}, 2427 (1991);
F.~Cooper, J.~M.~Eisenberg, Y.~Kluger, E.~Mottola and B.~Svetitsky,
Phys.\ Rev.\ D {\bf 48}, 190 (1993)
[hep-ph/9212206];
J.~M.~Eisenberg and G.~Kaelbermann,
Phys.\ Rev.\ D {\bf 37}, 1197 (1988);
T.~S.~Biro, H.~B.~Nielsen and J.~Knoll,
Nucl.\ Phys.\ B {\bf 245}, 449 (1984);
M.~Herrmann and J.~Knoll,
Phys.\ Lett.\ B {\bf 234}, 437 (1990),
D.~Boyanovsky, H.~J.~de Vega, R.~Holman, D.~S.~Lee and A.~Singh,
Phys.\ Rev.\ D {\bf 51}, 4419 (1995)
[hep-ph/9408214];
H.~Gies,
Phys.\ Rev.\ D {\bf 61}, 085021 (2000)
[hep-ph/9909500].
\bibitem{geiger}
K.~Geiger and J.~I.~Kapusta,
Phys.\ Rev.\ D {\bf 47}, 4905 (1993).
\bibitem{pcm}
K.~Geiger,
Phys.\ Rept.\  {\bf 258}, 237 (1995).
\bibitem{hosoya}
A.~Hosoya and K.~Kajantie,
Nucl.\ Phys.\ B {\bf 250}, 666 (1985).
\bibitem{zhang}
M.~Gyulassy, Y.~Pang and B.~Zhang,
Nucl.\ Phys.\ A {\bf 626}, 999 (1997)
[nucl-th/9709025];
B.~Zhang,
Comput.\ Phys.\ Commun.\  {\bf 109}, 193 (1998)
[nucl-th/9709009].
\bibitem{daniel}
P.~Danielewicz and M.~Gyulassy,
Phys.\ Rev.\ D {\bf 31}, 53 (1985).
\bibitem{weldon}
H.~A.~Weldon,
Phys.\ Rev.\ D {\bf 26}, 1394 (1982).

\bibitem{landshoff}
P.~V.~Landshoff and A.~Rebhan,
Nucl.\ Phys.\ B {\bf 383}, 607 (1992)
[Erratum-ibid.\ B {\bf 406}, 517 (1992)]
[hep-ph/9205235].
\bibitem{cutler}
R.~Cutler and D.~Sivers,
Phys.\ Rev.\ D {\bf 17}, 196 (1978).

\bibitem{combridge}
B.~L.~Combridge, J.~Kripfganz and J.~Ranft,
Phys.\ Lett.\ B {\bf 70}, 234 (1977).
\bibitem{biro}
T.~S.~Biro and B.~Muller,
Nucl.\ Phys.\ A {\bf 561}, 477 (1993)
[nucl-th/9211011].
\bibitem{heiselberg}
H.~Heiselberg and X.~Wang,
Nucl.\ Phys.\ B {\bf 462}, 389 (1996)
[hep-ph/9601247].

\end{thebibliography}
\end{document}